\documentstyle[aps,prb,preprint,tighten,floats,epsf]{revtex}
\newcommand{\deriv}[2]{\frac{d#1}{d#2}}

\begin{document}
\title{Decay of isolated surface features driven by
the Gibbs-Thomson effect in analytic model and simulation}

\author{James G. McLean,%
\thanks{Communicating author. Present Address:
Dept. of Chemistry, UCSD, La Jolla, California 92093-0358}
B. Krishnamachari,
D. R. Peale,%
\thanks{Present Address: Bell Laboratories, Lucent
Technologies, Murray Hill, NJ 07974}}
\address{Laboratory of Atomic and Solid State Physics,
Cornell University, Ithaca, New York 14853-2501}
\author{E. Chason,}
\address{Sandia National Laboratories, Albuquerque,
New Mexico 87185}
\author{James P. Sethna, and B. H. Cooper}
\address{Laboratory of Atomic and Solid State Physics,
Cornell University, Ithaca, New York 14853-2501}
\date{\today}
\maketitle

\begin{abstract}

A theory based on the thermodynamic Gibbs-Thomson relation
is presented which provides
the framework for understanding the
time evolution of isolated nanoscale
features (i.e., islands and pits) on surfaces.
Two limiting cases are predicted,
in which either diffusion or interface transfer
is the limiting process.
These cases correspond to similar regimes considered
in previous works
addressing the Ostwald ripening of ensembles of features.
A third possible limiting case is noted for the special geometry of
``stacked'' islands.
In these limiting cases,
isolated features are predicted to decay in size
with a power law scaling in time:
$A \propto (t_0-t)^n$, where $A$ is the area of the feature,
$t_0$ is the time at which the feature disappears,
and $n=2/3$ or 1.
The constant of proportionality is related to parameters
describing both the kinetic and equilibrium properties
of the surface.
A continuous time Monte Carlo simulation is used to test the
application of this theory to generic surfaces
with atomic scale features.
A new method is described to obtain macroscopic kinetic parameters
describing interfaces in such simulations.
Simulation and analytic theory are compared directly,
using measurements of the simulation to
determine the constants of the analytic theory.
Agreement between the two is very good
over a range of surface parameters,
suggesting that the analytic theory
properly captures the necessary physics.
It is anticipated that the simulation will be useful
in modeling complex
surface geometries often seen in experiments on physical surfaces,
for which application of the analytic model is not straightforward.

\end{abstract}

\narrowtext

\section{Introduction}
Real surfaces of crystalline materials
below their roughening temperature,
even of crystals with essentially perfect ordering,
are generally not in equilibrium.
More typically they have features such as irregular step edges,
excess individual adatoms or vacancies,
or clusters of adatoms or vacancies which we call islands and pits.
Since all these features have an associated free energy cost
but are not required by the macroscopic orientation of the surface,
their existence indicates that the surface
is not at a free energy minimum.
Such surfaces will therefore tend to relax towards equilibrium.

The way in which this relaxation takes place has been
a subject of interest for many years.
In processes which tend to produce nonequilibrium
features,
such as epitaxial growth or surface sputtering,
the relaxation may be important in determining
the time evolution of the
surface both during and subsequent to deposition or irradiation.
Surface relaxation will also be of particular interest
in the future for the field of
nanofabrication.
As artificially created structures approach the nanometer
scale, it becomes
important to understand whether these structures, once made,
will be relatively stable.
Stability of such features is crucial to success
for electronic interconnects or information recording applications.

One of the most studied situations is a surface populated
by an ensemble of islands of various sizes,
as might be found, for instance, in a growth
experiment after the initial stage of island nucleation.
This is a special case of a more general situation
in which small clusters of
a condensed phase of material, which are of dimensionality $D_c$,
exist in an environment of a vapor phase,
which is of dimensionality $D_v$.
Relaxation towards equilibrium in many such systems occurs through
the diffusion of atoms between the clusters, in a behavior
known as Ostwald ripening (also called coarsening).
Greenwood\cite{Greenwood56}
and Lifshitz and Slyozov\cite{Lifshitz58,Lifshitz61}
first considered the case $D_c = D_v = 3$
under the simplifying assumptions
that the inter-cluster mass flow is
limited by the diffusion process,
and that the total amount of the condensed phase is small.
Wagner\cite{Wagner61} extended the theory to include the
possibility that the transfer of adatoms between the phases
could be a limiting process.
Further extensions by Chakraverty\cite{Chakraverty67} and
others\cite{Wynblatt75,Thompson88,Dadyburjor86,Marqusee84}
applied the theory to the case of clusters on surfaces
(i.e., $D_v=2$, $D_c=2$ or 3).
References~\onlinecite{Zinke90} and~\onlinecite{Voorhees85}
include reviews of this material.

All of these treatments and many experimental
investigations\cite{Hanbucken84,Komura85,Zinke87,Shannon88,Zou88,%
Kruger89,Lagally89,Arrott90,Krichevsky93,Poirier95}
have concentrated on quantities,
such as average cluster radius and cluster density,
which are ensemble properties.
This is because the clusters involved are usually very small,
so that their properties are most readily studied by techniques,
such as diffraction, which sample large numbers of clusters.
With the advent of atomic scale microscopies
such as scanning tunneling microscopy (STM),
it has become practical to study the behavior of
individual surface features,
either within an ensemble or in isolation.
By observing the behavior of individual features,
the fundamental mechanisms responsible for stability and decay
can be studied without the complications which
result from the interactions
between the clusters of an ensemble system.
In addition, the behavior of an individual feature
may be the quantity of
interest in certain applications, as in nanofabrication.
Some studies have focused on the evolution of individual features,
either experimentally%
\cite{Trevor91,Peale92b,Cooper93,Michely93,Theis95,Morgenstern96}
or theoretically.\cite{Villain86,Dubson94}
For instance, we have found that,
on the Au(111) surface in air,
isolated islands decay in size such that the island area depends
roughly linearly on time.\cite{Peale92b}
A similar study of islands on the Ag(111) surface
in ultra high vacuum\cite{Morgenstern96}
observed island areas to be proportional
to time raised to an average exponent of $0.54$.

In this work we present an extension of the ensemble theories
to consider the decay of
isolated {\it individual\/} two dimensional islands on surfaces
($D_v=D_c=2$).
In Section~\ref{sec:theory} an analytic continuum model is presented
which accounts for three possible rate limiting steps
in island decay.
This model provides a connection between
the observed behavior of such features
and the parameters describing atomic processes on the surface.
Section~\ref{sec:limit} discusses limiting cases in this model
which are relevant to physical systems.
Specifically, we find ``interface-transfer limited'' and ``diffusion
limited'' cases
which are direct corollaries of
similar limits of Ostwald ripening.\cite{Chakraverty67,Zinke90}
Specific surface geometries lead to cases
which are unique to individual islands.
Islands completely surrounded by a step down or step up edge
can lead to
the ``outer-boundary limited'' and ``island in a pit'' cases.
Villain has considered the first of these geometries in
Ref.~\onlinecite{Villain86},
using assumptions which lead to the diffusion limited case.

Because the evolution of a single surface feature
involves a relatively
small area of the surface, it is feasible to perform
atomic scale simulation studies to test the analytic model.
Section~\ref{sec:sims} presents such a computer simulation study
of island decay for a generic surface with one particular geometry.
In Section~\ref{sec:macroparams} we discuss how to obtain
the parameters for the continuum model
which are appropriate to describing the atomistic simulation.
Using these parameters, we find in Section~\ref{sec:compare} that
comparison between the simulations and the analytic model
is very good,
showing that the analytic model captures the key physics
for describing feature evolution on a generic surface.
We anticipate that
this simulation technique will provide an additional tool
for analyzing the observed behavior of physical surfaces,
because it can be used to model complex
geometries which are often observed, e.g., in STM studies,
but which are not easily treated with the analytic model.

\section{Theory for decay of isolated surface features}
\label{sec:theory}
\subsection{The driving force for surface mass flow}
\label{sec:theory:gen}
At the heart of each of the treatments cited above is the
Gibbs-Thomson (GT) relation.
The pressure of a vapor which is in equilibrium with its condensed
phase depends on the curvature of the interface between the phases.
For a convex condensed phase the pressure is enhanced
relative to the pressure in equilibrium with a flat boundary.
The GT relation expresses this dependence.
The increased vapor pressure associated with a curved boundary is
the driving force behind the mass
transfer between different clusters or between a cluster
and a nearly straight step edge.

For a two dimensional, incompressible condensed phase,
the boundary of which has a radius of curvature $r$,
in contact with a two dimensional, noninteracting vapor,
the GT relation is\cite{GTisKelvin}
\begin{equation}
p_r^{\text{eq}}=p_\infty^{\text{eq}}\,
	\exp\!\left(\frac{\gamma\Omega}{rkT}\right)\;,
\label{eq:GT}
\end{equation}%
where $p_r^{\text{eq}}$ is the pressure at which the vapor will be
in equilibrium with the condensed phase,
$p_\infty^{\text{eq}}$ is the equilibrium vapor pressure
for a straight boundary ($r=\infty$),
$\gamma$ is the edge free energy (which is assumed to be
isotropic and independent of curvature),
and $\Omega$ is the area occupied by one atom
in the condensed phase.
When working with a lattice, it is more natural to consider
the vapor concentration $\rho$,
measured in units of $a^{-2}$ where $a$ is the lattice constant.
Because Eq.~(\ref{eq:GT}) assumes that the vapor is noninteracting,
the pressure will be proportional to the vapor concentration.
Thus we can substitute $\rho$ for $p$ in Eq.~(\ref{eq:GT}).

Note that the same relation continues to hold
for a condensed phase with a concave boundary.
In that case, the boundary is described
as having a negative radius of curvature.
For example, the adatom density inside a monolayer deep pit
will be reduced
relative to the pressure in equilibrium with a flat boundary.

If the vapor is interacting, as is the case for adatoms on a surface,
we have reported elsewhere\cite{Krishnamachari96}
that at vapor concentrations as low as
$\rho_r^{\text{eq}}\approx 10^{-3}\,a^{-2}$
there can be significant corrections to this equation.
However, by including a correction factor in the exponent
an equation of the same functional form as Eq.~(\ref{eq:GT})
continues to satisfactorily describe
the Gibbs-Thomson vapor enhancement in our system,
as will be discussed further in Sec.~\ref{sec:sims}.

The Gibbs-Thomson relation is sufficient
to anticipate the general behavior of a system.
Features with the highest curvature
will have the largest equilibrium adatom vapor concentrations.
On a surface with islands of different sizes as well as step edges,
a concentration gradient will be established
whereby adatoms diffuse away from high curvature
features and toward low curvature features.
For instance, large islands will grow as small
islands shrink and disappear.

\begin{figure}[b]
\centering\leavevmode
\epsfxsize=4in
\epsffile{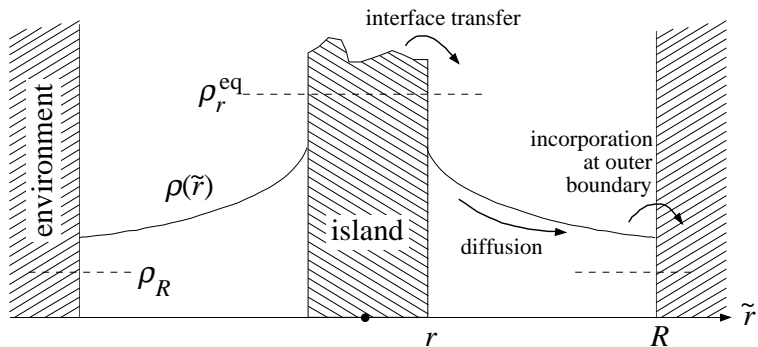}
\caption{A schematic representation of the adatom vapor density near a
small island of radius $r$ on a surface.
The equilibrium vapor densities for the island ($\rho_r^{\text{eq}}$)
and for the mean field environment at radius $R$ ($\rho_R$)
are indicated by dashed lines.
The actual density profile $\rho(\tilde{r})$ lies between these limits,
determined by the relative rates of the three processes of decay indicated
by arrows.}
\label{fig:profile}
\end{figure}

We now consider in more detail the specific case
of an isolated circular island on a surface
with an adatom concentration below the island's equilibrium value.
We shall use $r$ to denote the radius of the island,
and $\tilde{r}$ to
denote the radial distance from the center of the island.
The GT equation describes the equilibrium properties,
but in order to calculate the decay rate of an isolated island
we must consider the kinetics of the surface.
In the following sections, we will describe the three processes
involved in the flow of atoms from the island:
interface transfer,\cite{Wagner61,Chakraverty67}
diffusion,\cite{Lifshitz61}
and incorporation at the outer boundary.\cite{Marqusee84}
These three processes are illustrated in Fig.~\ref{fig:profile}.
We make the assumption throughout this treatment
that these processes are slow
on the time scale of local equilibration in the vapor,
so that the system may be described by a steady state solution
at any point in time.

\subsection{The interface transfer process}
\label{sec:theory:itproc}
As the island decays,
there must be a net detachment of atoms from the island.
If the deviation of the density of atoms just outside the island
$\rho(\tilde{r}\!=\!r)$
from the equilibrium density given by the GT relation
is sufficiently small,
then the current of atoms leaving the island $I_i$
will depend linearly on that deviation.
Further, if the curvature of the interface
does not significantly influence the manner
in which atoms leave the island
then the current will be proportional to the interface length,
so that
\begin{equation}
I_i=2\pi r K_i[\rho_r^{\text{eq}}-\rho(r)]\;,\label{eq:simpleIT}
\end{equation}
where we introduce the coefficient $K_i$
which is assumed to be independent of $r$.
The subscript `$i$' is chosen to identify the net detachment current
as the {\it interface transfer\/} current of atoms out of the island.

Further insight into the coefficient $K_i$ can be obtained
from a detailed balance argument.
In the remainder of this subsection we explore
how this coefficient arises from the microscopic dynamics,
thus clarifying the assumptions made above
and forming a basis for predicting the behavior of
physical systems in Sec.~\ref{sec:limit:phys}.
In order for an atom to attach to or detach from the island,
it must overcome some energy barrier at the island-vapor interface
(which depends on the particular microscopic atomic move).
Let $J_d$ be the detachment current density at the interface,
i.e., the rate per unit length of perimeter
at which atoms detach from the island edge.
Similarly, let $J_a$ denote the interface attachment current density.
Both of these will, in general, depend on
the interface curvature and the vapor concentration.
If we assume that both $\rho(r)$ and $[\rho_r^{\text{eq}}-\rho(r)]$
are $\ll 1\,a^{-2}$
[an assumption more stringent
than required for Eq.~(\ref{eq:simpleIT}),
but frequently satisfied]
then we may proceed by Taylor expanding
the current densities in $\rho(r)$,
obtaining
\begin{mathletters}
\begin{eqnarray}
J_d\bigl(r,\rho(r)\bigr)=&J_d^0(r)+&J_d'(r)\rho(r)
	+\frac{1}{2}J_d''(r)[\rho(r)]^2+\cdots\;,\\[0.1in]
J_a\bigl(r,\rho(r)\bigr)=&&J_a'(r)\rho(r)
	+\frac{1}{2}J_a''(r)[\rho(r)]^2+\cdots\;, \label{eq:Ja-exp}
\end{eqnarray}
\end{mathletters}%
where $J_d^0(r)$ is the detachment current density with no vapor,
and the prime indicates partial derivative with respect to $\rho$
evaluated at $\rho=0$.
Note that there is no constant term in Eq.~(\ref{eq:Ja-exp}) because
there can be no attachment unless an adatom vapor is present.
If the coefficients are well behaved
we can drop terms of higher than linear order.
Under equilibrium conditions,
when $\rho(r)=\rho_r^{\text{eq}}$,
the rate of detachment and attachment are equal,
\begin{equation}
J_d^0(r)+J_d'(r)\rho_r^{\text{eq}}
=J_a'(r)\rho_r^{\text{eq}}\;.\label{eq:det-bal}
\end{equation}
The total current $I_i$ is determined by the current densities,
which simplifies through Eq.~(\ref{eq:det-bal}),
\begin{eqnarray}
I_i&\equiv&2\pi r\,\left(J_d(r,\rho)-J_a(r,\rho)\right) \nonumber \\
&=&2\pi r\,\left(J_a'(r)-J_d'(r)\right)
	\left(\rho_r^{\text{eq}}-\rho(r)\right)\;.
\end{eqnarray}
Comparison with Eq.~(\ref{eq:simpleIT})
yields an expression for $K_i$,
\begin{equation}
K_i= J_a'(r)-J_d'(r)\approx J_a'(r)\;, \label{eq:Ki-from-J}
\end{equation}%
where we have used the fact that
in general we expect the detachment rate $J_d$
to have little dependence on the vapor concentration,
i.e., $|J_d'|\ll |J_a'|$.
Equation~(\ref{eq:Ki-from-J}) gives a clearer meaning to the
assumption that $K_i$ is independent of curvature;
it is equivalent to the assumption that $J_a'(r)$
(qualitatively the rate at which an atom will attach to the island
if it is already one step away,
i.e., a rate per adatom density)
is independent of $r$.
This assumption is plausible;
for instance, if we have an ideal gas in an enclosure,
the number of gas atoms hitting a given area of the enclosure wall is
independent of its curvature.
However, this assumption should be justified for any particular
system under study.

\subsection{The diffusion process}
\label{sec:theory:diff}
Once the atoms have detached from a decaying island,
they must diffuse away across the terrace.
Under our assumption of quasi-steady state,
the rate at which they diffuse away is governed by
the two-dimensional (2D) time-independent radial diffusion equation.
The general solution of this equation
diverges logarithmically at large distances.
It is therefore necessary to specify a boundary condition
far from the island.
It is assumed that the island is sufficiently isolated
from other surface features
so that a mean field approach is valid.
Rather than specify the island's environment in detail,
we apply a generic boundary condition that
the vapor concentration is some $\rho(\tilde{r}\!=\!R)$
at a radius $R>r$,
where $R$ may vary in time.
We will consider some particular examples below.
The current of adatoms is then given by the usual solution to the
steady state 2D radial diffusion equation:
\begin{mathletters}
\begin{eqnarray}
I_D&=&2\pi\frac{D}{\ln(R/r)}(\rho(r)-\rho(R)) \nonumber \\
   &=&2\pi K_D(\rho(r)-\rho(R))\label{eq:diffcurr}
\end{eqnarray}
\begin{equation}
K_D\equiv\frac{D}{\ln(R/r)}\;,
\label{eq:KDdef}
\end{equation}%
\end{mathletters}%
where we introduce the coefficient $K_D$.
$D$ is the collective (or chemical) diffusion coefficient
of the vapor on a step-free terrace.
The subscript `$D$' is chosen to denote {\it diffusion\/}.

\subsection{The process of incorporation at the outer boundary}
\label{sec:theory:incorp}
Finally, the adatoms must leave the vapor
by crossing the outer boundary,
thus becoming incorporated into the mean field.
Since we have left unspecified the nature of the outer boundary,
this process may limit the rate at which atoms can leave.
We may reason in a manner similar to the case of interface transfer,
as discussed above.
There will be some concentration $\rho_R$ such that
if $\rho(R)=\rho_R$ then the net current
across the outer boundary vanishes.
If $\rho(R)$ does not vary much from $\rho_R$,
then the current across the outer boundary will be
proportional to their difference,
\begin{equation}
I_b=2\pi R K_b(\rho(R)-\rho_R)\;.\label{eq:OBcurr}
\end{equation}
This equation introduces the rate coefficient $K_b$,
where the subscript `$b$' is chosen to denote the outer
{\it boundary\/}.
Note the similarity between Eqs.~(\ref{eq:OBcurr})
and~(\ref{eq:simpleIT}).

\subsection{The total mass flow and island decay rate}
As commented at the end of Sec.~\ref{sec:theory:gen},
we assume that the system is in a steady state.
Therefore the three atom currents, which are in series, must be
equal.
Equating Eqs.~(\ref{eq:simpleIT}), (\ref{eq:diffcurr}),
and~(\ref{eq:OBcurr})
gives the following expressions
for the concentrations of adatoms at the boundaries
and for the net detachment current
in terms of equilibrium concentration values:
\begin{mathletters}
\begin{equation}
\rho(\tilde{r}\!=\!r)=\left(\frac{1}{K_bR}
		+\frac{1}{K_D}\right)C(r)\rho_r^{\text{eq}}
	+\frac{1}{K_ir}C(r)\rho_R\;,
\end{equation}
\begin{equation}
\rho(\tilde{r}\!=\!R)=\frac{1}{K_bR}C(r)\rho_r^{\text{eq}}
	+\left(\frac{1}{K_D} +\frac{1}{K_ir}\right)C(r)\rho_R\;,
\end{equation}
\end{mathletters}
\begin{mathletters}
\label{eqs:totcurr}
\begin{eqnarray}
I&=&2\pi C(r)\:(\rho_r^{\text{eq}}-\rho_R)\;, \\
 &=&2\pi C(r)\,\rho_\infty^{\text{eq}}\left[
	\exp\!\left(\frac{\gamma\Omega}{rkT}\right)
	-\frac{\rho_R}{\rho_\infty^{\text{eq}}}\right]\;,
\label{eq:totcurrexp}
\end{eqnarray}%
\end{mathletters}%
where
\begin{equation}
C(r)\equiv\left(\frac{1}{K_ir}+\frac{1}{K_D}
			+\frac{1}{K_bR}\right)^{-1}\;.\label{eq:Cdef}
\end{equation}%
In the Eq.~(\ref{eq:totcurrexp}),
the GT relation has been substituted for $\rho_r^{\text{eq}}$.
In Eq.~(\ref{eq:Cdef}) the coefficients
of the individual current equations
add together as conductances.
Thus $C(r)$ is the overall conductance relating the adatom current
to the driving force provided by an imbalance
in equilibrium densities.
Note that $\rho(r)$ and $\rho(R)$ are intermediate
between the equilibrium
concentrations $\rho_r^{\text{eq}}$ and $\rho_R$
(see Fig. \ref{fig:profile}).
In this way a driving force is present for all three processes
in the mass flow away from the island.

\section{Limiting cases of the analytic model}
\label{sec:limit}
\subsection{Simplifying mathematical limits}
\label{sec:limit:math}
The form of Eqs.~(\ref{eqs:totcurr}) and~(\ref{eq:Cdef})
is quite complex, especially in their dependence on $r$.
In order to better understand them
it is useful to consider simplifying mathematical limits.
Using physically reasonable approximations,
we will find limits
in which the decaying island size obeys a simple power law.
That is, in these limits the area of the island obeys
$A \propto (t_0-t)^n$
where $A$ is the area of the island
and $t_0$ is the time at which the island disappears.
It will be seen that
the exponent $n$ may take the values of $2/3$ or 1.

Two approximations must be satisfied in order to achieve
either of the power law limits.
The first approximation is that
(i)~the equilibrium adatom density at the outer boundary is
close to the equilibrium density for a straight step edge,
$\rho_R\approx \rho_\infty^{\text{eq}}$.
We expect this to be valid if the island edge has a much higher
curvature than other nearby surface features, so that
$|\rho_r^{\text{eq}}-\rho_\infty^{\text{eq}}|\gg
|\rho_R-\rho_\infty^{\text{eq}}|$.
We will thus label this as the ``$R\gg r$'' approximation.

The second approximation required is that
(ii)~the argument of the exponential
in Eq.~(\ref{eq:totcurrexp}) must be much less than one,
so that the exponential may be approximated
by the constant and linear terms from its Taylor expansion.
This approximation we will label as the ``GT expansion''.
It is also frequently physically reasonable.
For instance, for a surface with $\gamma=0.117\,\text{eV}/a$
(see Sec.~\ref{sec:sims}) at room temperature,
approximation (ii) is good to within 10\% for islands of size
$r\agt 23\,a$.
This approximation is not always valid, however;
for instance, it seems to fail for island decay observed
by Morgenstern {\it et~al.}\cite{Morgenstern96}

Using these first two approximations,
Eq.~(\ref{eq:totcurrexp}) may be rewritten
\begin{equation}
I \approx 2\pi C(r)\,\rho_\infty^{\text{eq}}
		\frac{\gamma\Omega}{rkT}\;.\label{eq:expapx}
\end{equation}
In order to get a power law decay of the island size,
a third required approximation is that (iii)~$C(r)$ must be constant
or proportional to a power of $r$.
This can be realized in two ways.
(Physical realizations of the following approximations
will be discussed in Sec.~\ref{sec:limit:phys}.)

In one limit (iii-a) $C(r)$ may be approximated
to be a constant $C_0$.
This can occur, for instance,
if the third term of Eq.~(\ref{eq:Cdef})
is much larger than the first two,
i.e., $K_bR \ll \min(K_ir$, $K_D)$,
and if R is constant in time.
This we call the ``constant $C$'' approximation.
The adatom current gives the time rate of change for the island area,
so that in this case we have
\begin{mathletters}
\label{eqs:all1/3}
\begin{equation}
2\pi r\deriv{r}{t}=\deriv{A}{t}= -I\Omega=
	-2\pi\Omega\,C_0\,\rho_\infty^{\text{eq}}
	\frac{\gamma\Omega}{rkT}\;.
\end{equation}
This may be directly integrated to obtain
\begin{equation}
r = \left(3C_0\frac{\rho_\infty^{\text{eq}}\,\gamma\Omega^2}
		{kT}\right)^{1/3}
	(t_0-t)^{1/3}\;,\label{eq:1/3decay}
\end{equation}
\begin{equation}
A\propto(t_0-t)^n\;,\qquad n=2/3\;.
\end{equation}
\end{mathletters}%
This will be referred to as the ``$n=2/3$ mathematical limit''.

In the other limit (iii-b) of Eq.~\ref{eq:Cdef}
which can lead to power law behavior,
$C(r)$ is approximately proportional to $r$.
This we call the ``linear $C$'' approximation.
This can occur, for instance,
if the first term of Eq.~(\ref{eq:Cdef})
is much larger than the last two.
Expressing the $r$ dependence of $C(r)$ explicitly with $C(r)=C_1r$,
analogously to Eqs.~(\ref{eqs:all1/3}) we have
\begin{mathletters}
\label{eqs:all1/2}
\begin{equation}
2\pi r\frac{dr}{dt}=\frac{dA}{dt}= -I\Omega=
	-2\pi\Omega\,C_1\,\rho_\infty^{\text{eq}}
	\frac{\gamma\Omega}{kT}\;,
\end{equation}
\begin{equation}
r = \left(2C_1\frac{\rho_\infty^{\text{eq}}\,\gamma\Omega^2}
			{kT}\right)^{1/2}
	(t_0-t)^{1/2}\;,\label{eq:1/2decay}
\end{equation}
\begin{equation}
A\propto(t_0-t)^n\;,\qquad n=1\;.
\end{equation}
\end{mathletters}
In this ``$n=1$ mathematical limit'',
the island area scales linearly with time.

Note that in both of these limits,
the velocity of the interface $dr/dt$ diverges
as the island approaches the vanishing point at time $t_0$.
If a high enough velocity is reached,
then the assumption that the system is in a quasi-steady state
will fail.
In practice, the velocity is limited by the fact
that the island cannot be smaller than one atom,
so that the breakdown of quasi-steady state might not occur.

The formalism presented above for describing the decay of islands
via adatom diffusion
may also be applied to the decay of pits in the surface
(vacancy clusters) via vacancy diffusion.
These pits may be the source of vacancies
which diffuse in the surface and
recombine at step edges.
In principle, both adatom and vacancy motion
can contribute to mass flow on the surface.
However, the rate constants may be very different for
adatom and vacancy motion,
so that one may dominate the surface evolution.

\subsection{Particular examples for physical systems}
\label{sec:limit:phys}
In this section we will discuss limiting behaviors
associated with each of
the three processes involved in the mass flow from a decaying island,
i.e. interface transfer, diffusion, and outer boundary incorporation
(see Secs.~\ref{sec:theory:itproc}, \ref{sec:theory:diff},
and~\ref{sec:theory:incorp} respectively).
The special case of an island located within a pit
will also be discussed.

The treatment of Sec.~\ref{sec:theory} bears similarities
to treatments concerned with Ostwald ripening
in ensembles of clusters on surfaces.\cite{Chakraverty67,Zinke90}
In those treatments, a central quantity of interest
is the critical radius $r_c$,
defined such that islands of size $r<r_c$ shrink
while larger islands grow.
Under approximations analogous to those in Sec.~\ref{sec:limit:math},
along with certain assumptions about the ensemble size distribution,
the critical radius is found to increase in time
according to a power law.\cite{Zinke90}
Indeed, it can be shown that the resulting exponents are identical
to those found above for
an individual shrinking island.\cite{McLean96}
Note, however, that the critical radius of an ensemble {\em grows},
while the physical radius of an isolated island {\em shrinks}.

The physical characteristics which lead to power law behavior
in Ostwald ripening\cite{Chakraverty67,Zinke90}
will also lead to corresponding behavior for isolated islands.
Consider a configuration in which an island is not
surrounded by any clearly defined outer boundary.
For instance, it might be on a terrace bounded on two sides
by relatively straight steps.
We assume that the outer boundary incorporation process
described in Sec.~\ref{sec:theory:incorp} does not significantly
affect the island decay,
so that the third term in Eq.~(\ref{eq:Cdef}) is negligible.
If the curvature of the island edge is higher
than that of other surface features,
i.e., if the island is small,
then the approximation (i)($R\gg r$) remains valid.
The island must be larger than than the length scale
set by the step free energy,
so that the approximation (ii)(GT expansion)
is also valid.
The choice of a mean field outer boundary $R$ is problematic
(it still enters the problem through $K_D$).
If it is very difficult for atoms to cross
from the condensed phase to
the vapor phase and vice versa at the island edge,
then the first term of Eq.~(\ref{eq:Cdef}) dominates the others.
In this way $C(r)\approx K_ir$,
so that the outer boundary is totally removed from the problem.
The approximation (iii-b)(linear $C$)
and Eqs.~(\ref{eqs:all1/2}) apply.
This is called the ``interface-transfer limited'' case,
and falls into the $n=1$ mathematical limit.

If, on the other hand, the isolated island is in a system where
diffusion is very slow,
then we have the ``diffusion limited'' case,
for which the second term of Eq.~(\ref{eq:Cdef}) dominates.
Since $C(r)\approx K_D$ depends on the outer boundary
through Eq.~(\ref{eq:KDdef}),
further assumptions must be made.
The classic treatment of this regime\cite{Chakraverty67}
assumes that the weak logarithmic dependence of $K_D$ on $r$ and $R$
may be neglected.
This is equivalent to assuming that there is
a vapor phase screening length $\xi=(\lambda-1)r$.
This screening length leads to an outer boundary that goes as
$R=\lambda r$,
so that $C(r)\approx K_D=D/\ln(\lambda)$ is constant.
Thus the approximation (iii-a)(constant $C$)
and Eqs.~(\ref{eqs:all1/3}) apply,
and the diffusion limited case falls
into the $n=2/3$ mathematical limit.
It should be noted, however,
that the screening length hypothesis
is not physically self-consistent,
so that this approach should not be expected
to yield quantitatively correct results.
In particular, a screening length can only arise
through the interaction of many features
(islands, pits, and step edges),
in which case it turns out
that the screening length is too long
to properly treat the island as isolated.\cite{Marqusee84,McLean96}

A cross over between the interface-transfer limited
and diffusion limited cases can be defined by the condition
that the first two terms of Eq.~(\ref{eq:Cdef}) are equal.
At sufficiently large radii $K_ir$ will be larger than $K_D$,
and the mass flow {\it must\/} be in the diffusion limited regime.
Only when the island reaches a sufficiently small size
can the interface-transfer limited regime be observed.
Physically, a shorter island perimeter
provides fewer detachment sites,
making interface transfer more difficult.
We can define a transition radius
at which this cross over will occur by
equating $K_i r$ and $K_D$, which gives
[see Eqs.~(\ref{eq:Ki-from-J}) and~(\ref{eq:KDdef})]
\begin{equation}
r_t=\frac{K_D}{K_i}
=\left(\frac{D}{\ln(\lambda)J_a'}\right)\;.
\label{eq:rtrans}
\end{equation}
Since $\ln(\lambda)$ is a number of order one,
we see that in order for $r_t$ to be significant,
$J_a'$ must be significantly larger than $D$.
That is,
there must be some impediment to an atom attaching to an island
relative to diffusion.
On a clean metal surface this is unlikely,
as attachment energy barriers are generally found to be similar to
or lower than diffusion barriers.\cite{Liu93,Scheffler94}
For adsorbate covered surfaces, however,
the adsorbate will often tend to bind to the step edge
and may therefore present a steric impediment to attachment.

For individual islands, other cases arise which have no analog
in the ensemble treatments.
For instance, a configuration frequently observed on surfaces
(e.g., in growth or ion bombardment)
is one where the decaying island is on top of another larger island,
forming a multilayer structure.
Focusing on the decay of the top layer,
the outer boundary of our treatment
is the perimeter of the second layer.
Because the energy barrier for interlayer transport of atoms
is frequently high,
the rate coefficient $K_b$ may be very small
for this multilayer case.
This situation, which we call the ``outer-boundary limited'' case,
satisfies approximation (iii-a)(constant $C$).
In this case the equilibrium density at the outer boundary is
$\rho_R=\rho_R^{\text{eq}}$ as given by the Gibbs-Thomson relation.
However, if the lower island is much larger than the upper one
then the approximation (i)($R\gg r$) is also reasonable,
so that $\rho_R\approx\rho_\infty$ and
Eqs.~(\ref{eqs:all1/3}) apply with $C=K_b R$.
Thus the outer-boundary limited case falls
into the $n=2/3$ mathematical limit.
This configuration was considered by Villain,\cite{Villain86}
who also found an exponent of $n=2/3$.
Note however that Villain's approximations
lead to the diffusion limited case discussed above.
Our outer-boundary limited case gives the same exponent,
but through a different mechanism.

A final surface configuration of interest
is the ``island in a pit'' case,
where an island is within a monolayer pit.
The outer boundary is well defined,
similarly to the previous case.
Here, however, the atomic processes at the outer boundary
are the same as those taking place at the inner boundary,
so that $K_b=K_i$.
There is no mathematical power law limit for this case,
as neither approximation (iii-a) nor (iii-b) apply.
We can, however, give a criterion as to whether the decay is
dominated by the diffusion process or the interface transfer process.
This criterion is similar to the cross over
between the interface-transfer limited
and diffusion limited cases above.
This transition radius $r'_t$ is defined
by the condition that the second term of Eq.~(\ref{eq:Cdef}) is
equal to the sum of the first and third,
yielding the transcendental equation
\begin{equation}
\left(\frac{R}{r'_t}+1\right)^{-1} \ln(R/r'_t) = \frac{D}{K_iR}\;.
\label{eq:rtranspit}
\end{equation}
The left hand side of Eq.~(\ref{eq:rtranspit})
has a maximum value of approximately 0.278;
if the right hand side is larger than this,
then there is no transition
radius and the entire decay is interface-transfer limited.
Otherwise, the decay is interface-transfer limited
for very large and very
small islands, but diffusion limited between.
The physical interpretation is that
when the island is almost as large as the pit,
diffusion of atoms across the narrow terrace
between them occurs rapidly,
leading to interface-transfer limited behavior for large islands.
For very small islands, the short island perimeter leads to
interface-transfer limited decay through lack of detachment sites.

Finally, a comment on how experimentally controllable parameters
should be expected to influence feature behavior.
The overall rate of decay
[see the prefactors of Eqs.~(\ref{eq:1/3decay})
and~(\ref{eq:1/2decay})]
depends on many parameters, and therefore could
vary widely from material to material.
For any particular material,
temperature will play a central role in controlling the decay rate,
primarily through the temperature dependence
of $\rho_\infty^{\text{eq}}$.
This is because the equilibrium density results
from adatoms being thermally
activated to detach from step edges,
and therefore will depend exponentially on inverse temperature.
The effect adsorbates would have on a system is
another particularly interesting influence on the decay
rate.\cite{Trevor89,Peale92b}
Adsorbates lower the free energy of a surface.
If they also lower the relative free energy of individual substrate
adatoms,
this could substantially increase $\rho_\infty^{\text{eq}}$,
leading to an increase in the decay rate.
Adsorbates would also be expected to lower the step free energy,
improving the approximation (ii)(GT expansion)
[see Eq.~(\ref{eq:expapx})].
However, if the adsorbates interact with the surface
chemically through directional bonding,
then such simple arguments are insufficient;
other considerations, outside the scope of this paper,
must be included.
Finally, adsorbates could also affect rate constants
by simply getting in the way of the substrate atoms,
as mentioned above in connection with Eq.~(\ref{eq:rtrans}).

\section{Computer simulations of island decay}
\label{sec:sims}
The continuum model presented above is useful
for identifying important
factors controlling the stability and decay of surface features.
However, it makes many simplifying assumptions,
in particular about the proper macroscopic description of microscopic
behavior.
To approach the problem from another direction,
we may examine island decay for a generic system based on simple
energetics,
where different microscopic processes are explicitly included.
The results may then be compared to the thermodynamic model
to see how well the latter captures the behavior of the system.
To this end we have pursued computer simulations of island decay.
One goal of these simulations is
to verify that the above continuum treatment
is applicable to nanoscale systems based on simple energetics,
and in particular to test whether
the limiting regimes of decay predicted by the thermodynamic model
can be reproduced in simulations.
We also hope to illuminate what microscopic mechanisms
play the dominant role in the mass transfer process,
and how they relate to the thermodynamic parameters.

Our computer code simulates atoms moving on a square lattice
using the solid-on-solid model, which does not allow overhangs.
The system is rectangular with periodic boundary conditions.
Atomic moves are classified according to the initial
horizontal coordination of the
moving atom (0 to 4 for the square lattice),
the coordination the atom would have if the move were made,
and whether the vertical motion is up, down, or absent.
Taking into consideration that some moves
are geometrically disallowed,
this yields 48 types of moves.
The rates for these move types are determined by an Arrhenius form,
in which a single attempt frequency $\nu$ and energy barriers
for each type of move are parameters in the simulation.
Note that the attempt frequency applies to each move separately,
so that each atom attempts to move at the rate $4\nu$.
In order to ensure that detailed balance is satisfied,
these barriers are
constrained such that the system obeys the bond counting Hamiltonian
\begin{equation}
{\cal H} = -\,\frac{1}{2}\sum_{\text{sites}}\varepsilon_B
\times(\text{coordination})\;,
\end{equation}
where $\varepsilon_B$ is the energy of a single bond.
This constraint leads to 26 independent barriers, along with the bond
energy and attempt frequency, as parameters defining the system.
A dynamical Monte Carlo algorithm\cite{Bortz75,Voter86,Chason91}
is used in which every computer
iteration results in an atomic move,
the simulated time being advanced probabilistically.
This algorithm is very efficient relative to
standard Monte Carlo methods,
as no attempted moves are rejected.

Although we are studying the generic behavior of surfaces,
we wish to have barriers
which are reasonable from a physical standpoint.
Therefore, the energy barriers for in-plane moves used in the
following simulations are based on barriers for the Cu(100) surface
as calculated with
the Finnis-Sinclair atom embedding model.\cite{Finnis84}
In Ref.~\onlinecite{Breeman94} barriers were calculated
for in-plane moves with all $2^{10}$ possible configurations
of nearest and next-nearest neighbors to
the initial and final sites of the moving atom.
Exchange type moves were considered,
but were found to be unimportant for this surface.
For each combination of initial and final coordination
we averaged an appropriate subset of these $2^{10}$ barriers
to obtain the barriers for use in our simulation.
This procedure clearly does not retain the details
of the Cu(100) surface.
However it does yield physically reasonable numbers,
and in particular results in a bond energy
consistent with the original barrier set.
The energy barriers we use,
corresponding to a bond energy of $\varepsilon_B=0.341\,\text{eV}$,
are shown in Table~\ref{tab:barriers}.

\mediumtext
\begin{table}[thb]
\caption{Energy barriers for intra-layer atomic moves, in units of eV}
\label{tab:barriers}
\begin{tabular}{cdddd}
\bf{Initial}      &\multicolumn{4}{c}{\bf Final Coordination} \\ \cline{2-5}
\bf{Coordination} & 0-fold   & 1-fold   & 2-fold   & 3-fold   \\ \hline
0-fold            & 0.697    & 0.479    & 0.328    & 0.166    \\
1-fold            & 0.820    & 0.624    & 0.450    & 0.275    \\
2-fold            & 1.010    & 0.791    & 0.591    & 0.377    \\
3-fold            & 1.189    & 0.957    & 0.718    & 0.462
\end{tabular}
\end{table}
\narrowtext

The barriers for interlayer movement were set very high
($\approx 100\,\text{eV}$) in order to prevent
the occurrence of such moves.
In order for the decay to take place within a reasonable amount of
computer time,
the simulated temperature was a moderately high 1347\,K,
or 60\% of the critical temperature for this model.
Note that the 2D critical temperature is a lower bound for the
roughening temperature
of the three-dimensional system.\cite{Beijeren75}
The attempt frequency used was
$\nu=10^{12}\,\text{s}^{-1}$, which sets the overall time scale.

Elsewhere we have shown\cite{Krishnamachari96} that this model
does exhibit a Gibbs-Thomson effect.
That is, smaller islands are in equilibrium
with higher adatom concentrations.
There we show that the parameters of Eq.~(\ref{eq:GT}) can be
analytically calculated\cite{Krishnamachari96,Avron82}
for this simulation to be 
$\rho_\infty^{\text{eq}}=3.58\times\!10^{-3}\,a^{-2}$
and $\gamma=0.117\,\text{eV}/a$.
However,
at the simulation temperature of 1347\,K,
the adatom concentration of the vapor phase is high enough so
that its equation of state deviates significantly from that of a
noninteracting gas.
Therefore, the proper Gibbs-Thomson relation
is much more complex than Eq.~(\ref{eq:GT}).
Nevertheless, we found\cite{Krishnamachari96} phenomenologically
for this model at these temperatures
that the GT vapor enhancement can be well described by including
a correction factor $g=1.59$
in the exponent of a {\it modified} Gibbs-Thomson equation
\begin{equation}
\rho_r^{\text{eq}}=\rho_\infty^{\text{eq}}\,
	\exp\!\left(\frac{g\gamma\Omega}{rkT}\right)\;.
\end{equation}

\begin{figure}[b]
\begin{center}
\begin{tabular}{ccc}
\epsffile{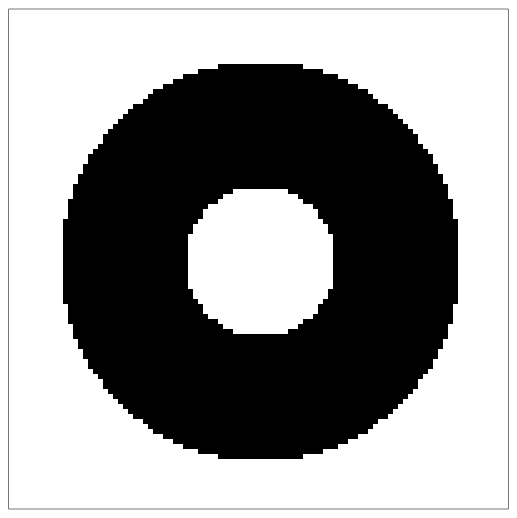} &\epsffile{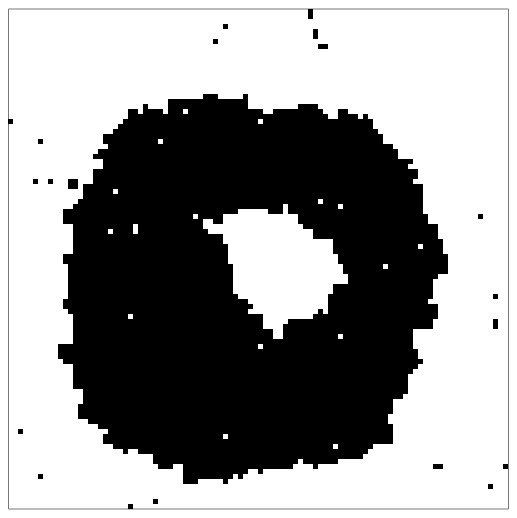}
	&\epsffile{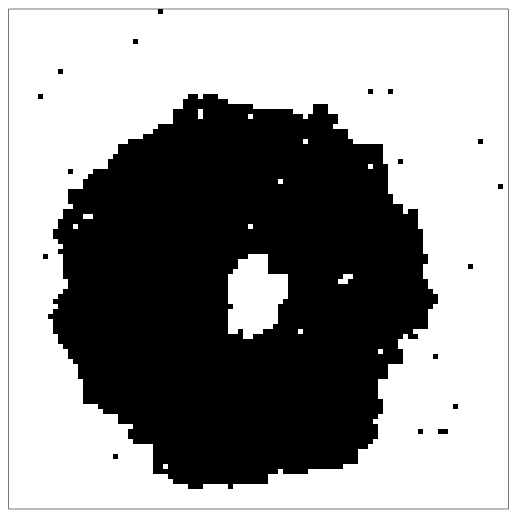} \\
(a) $t=0.0\,$s	&(b) $t=0.5\times\!10^{-4}\,$s	&(c) $t=1.0\times\!10^{-4}\,$s
\end{tabular}
\end{center}
\caption{Snapshots of a simulated island decay:
the initial configuration (a)
and later stages at (b)~45\% and (c)~90\% of the total decay time.
White regions are one atom higher than black regions.}
\label{fig:simpicts}
\end{figure}

\begin{figure}[b]
\centering\leavevmode
\epsfxsize=4in
\epsffile{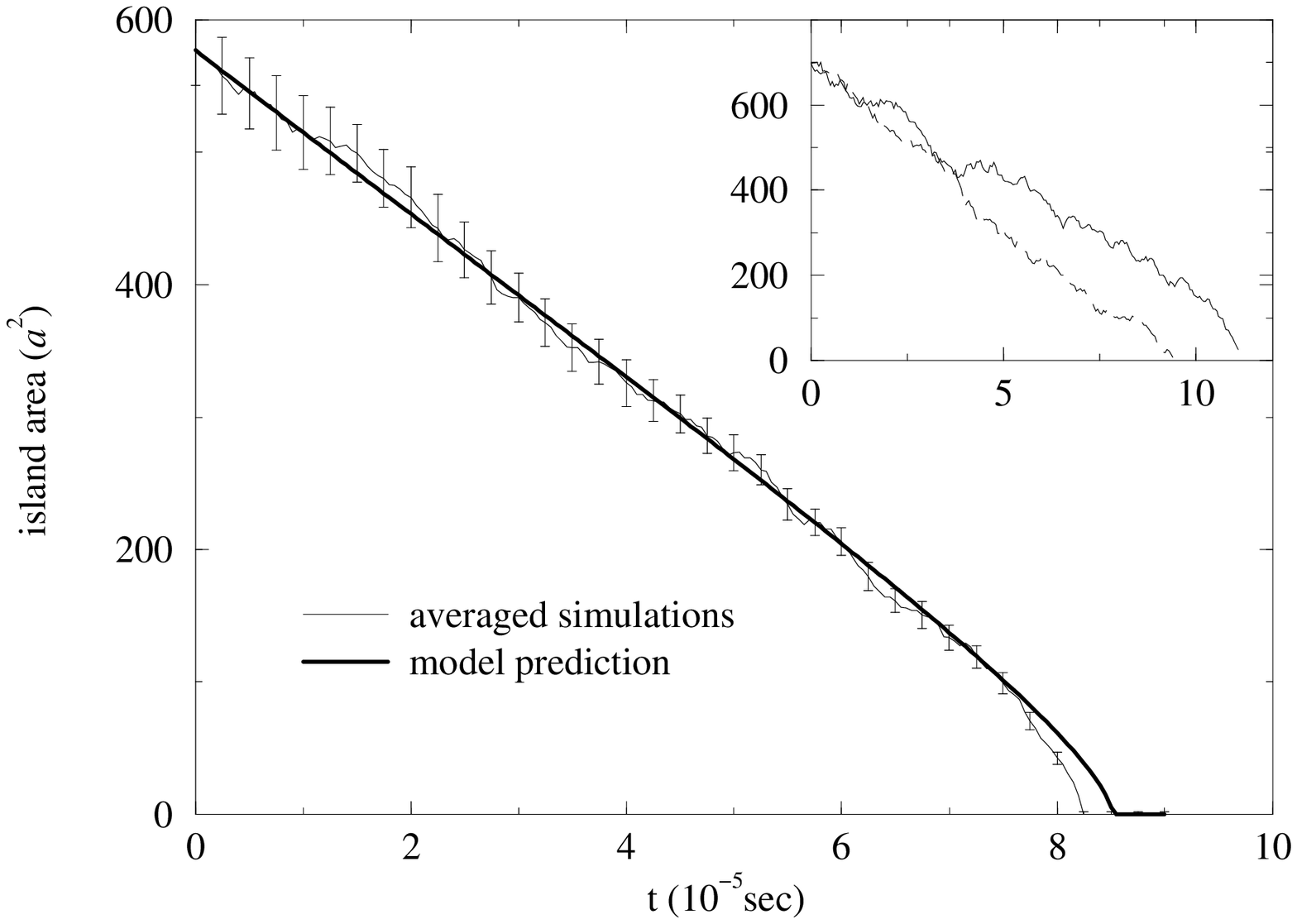}
\caption{Simulated island decay using barriers based on Cu(001),
showing the island area versus time.
The figure shows the average area of ten simulations
referenced to time of island disappearance.
Error bars represent standard deviations of the mean.
The inset shows two representative individual decays.}
\label{fig:Cu-decay}
\end{figure}

For our decay simulations, a $100\,a\times 100\,a$ lattice was used.
The initial configuration was a circular island,
$15\,a$ in radius, centered within a pit of radius $40\,a$
[as in Fig.~\ref{fig:simpicts}(a)].
As expected for a system exhibiting a Gibbs-Thomson effect,
the island disappears over time
as the adatoms move from the island to the pit edge.
Figure~\ref{fig:simpicts} shows configurations
typical of those observed during decay;
the island remains relatively circular,
and remains relatively well centered in the pit.
In any individual simulation,
the area of the island showed significant
fluctuations during the decay (see Fig.~\ref{fig:Cu-decay} inset).
To obtain an averaged behavior,
ten simulations were run with identical initial conditions
except for the seed for the random number generator.
Each decay represents roughly 15 hours of CPU time
on an IBM RS6000 model 550 computer.
The time origin of each run was offset
so that the islands vanished at the same time,
and the sizes of the islands were averaged at each sampled time.
Very early times, during which the system was
coming into steady state, were ignored.
Figure~\ref{fig:Cu-decay} shows the averaged results.

We wish to make quantitative comparisons between these results
and the macroscopic theory of Sec.~\ref{sec:theory}.
In particular, we are simulating the ``island in a pit'' case
discussed in Sec.~\ref{sec:limit:phys}.
As discussed there, the outer boundary rate coefficient is $K_b=K_i$.
Because the geometry of the system is completely specified,
it is unnecessary to make assumptions about screening lengths.
The outer boundary position $R$ is very close to constant,
as the area of the island is small compared to the area of the pit.
Finally, the equilibrium density at the outer edge is
given by the GT relation as $\rho_R=\rho_{-R}^{\text{eq}}$.
Note that the pit edge has a negative radius of curvature,
which changes the sign of the exponent in the GT relation.
Thus the equation governing the decay of the island in this system is
[see Eqs.~(\ref{eqs:totcurr})]
\begin{mathletters}
\begin{equation}
\deriv{r}{t}=
-\frac{\Omega}{r}\left(\frac{1}{K_D}+\frac{1}{K_i}
				\left[\frac{1}{r}+\frac{1}{R}\right]
		 \right)^{-1} \rho_\infty^{\text{eq}}
\left( \exp\left(\frac{g\gamma\Omega}{rkT}\right)
	- \exp\left(-\frac{g\gamma\Omega}{RkT}\right) \right)
\;,\label{eq:simcurr}
\end{equation}
\begin{equation}
K_D\equiv\frac{D}{\ln(R/r)}\;.
\end{equation}
\end{mathletters}%
Here we have included
the phenomenological correction factor $g$
in the modified GT expression.
The definition of $K_D$ is explicitly repeated
to emphasize that it has significant dependence on $r$
for this configuration.

Note that neither the approximation (i)($R\gg r$)
nor (ii)(GT expansion) applies to these simulations.
Specifically, the island is too large relative to the pit
for~(i) to be satisfied,
and the island is too small for~(ii) to be satisfied.
Therefore, the complete decay equation above must be used
for a quantitative comparison of simulation and analytic theory.

\section{Determination of macroscopic parameters
for the microscopic simulation}
\label{sec:macroparams}
In order to apply the macroscopic decay equation to the simulations,
it is necessary to determine the correct values of the parameters
$D$ and $K_i$.
These may be determined through proper measurements of the simulation
behavior
together with analytic thermodynamic calculations.
Portions of the following treatment are achieved through the use of
a mapping of the lattice gas to the Ising model,
for which a high field (low concentration) expansion is used
to obtain the equation of state.
The interested reader can find a more complete description
of this mapping and its use in Ref.~\onlinecite{Krishnamachari96}.

\begin{figure}[b]
\centering\leavevmode
\epsfxsize=4in
\epsffile{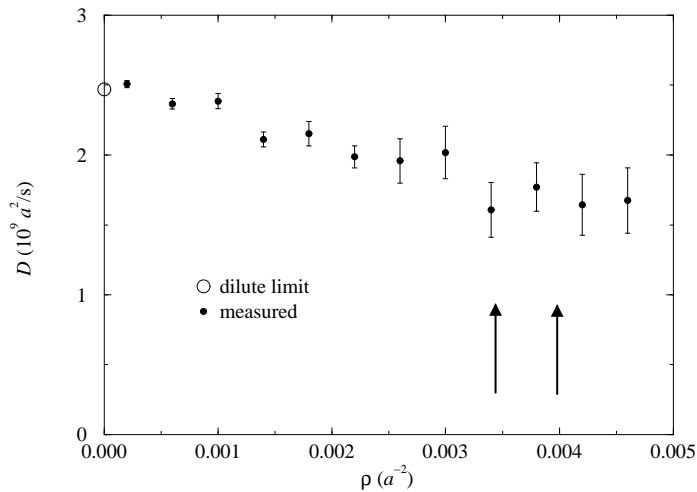}
\caption{The dependence of the diffusion coefficient on adatom density in
the simulation,
obtained from direct measurements
of current densities under enforced concentration gradients.
Open circle shows the theoretical dilute limit value.
The arrows indicate the concentration of vapor adatoms required
for equilibrium with
each of the pit (left arrow) and the island (right arrow)
at the beginning of the decay simulation.}
\label{fig:CuD}
\end{figure}

In the limit of vanishing concentration,
the diffusion coefficient can be directly determined from the attempt
frequency and the energy barrier
for zero initial and final coordination:
\begin{equation}
D(\rho \to 0)=a^2\nu e^{-E_{00}/kT}=2.47\times\!10^9\,a^2/\text{s}\;.
\end{equation}
However, as noted above,
the temperature of the simulation results in an
adatom concentration which is high enough
for the system to deviate significantly from ideal gas behavior.
Therefore the diffusion coefficient was measured
directly with ``enforced gradient'' simulations.
Simulations were run in which a flat terrace populated by adatoms
was divided into three sections.
The densities in the left and right sections were held constant
at values differing by $4\times10^{-4}\,a^{-2}$
(a small difference compared with
$\rho_\infty^{\text{eq}}=3.58\times\!10^{-3}\,a^{-2}$).
The resulting adatom current density was then measured
and used to calculate a diffusion coefficient
through Fick's Law,
$J = -D \nabla \rho$ where $J$ is the adatom current density.
Figure~\ref{fig:CuD} shows the measured $D$
versus the average of the densities in the left and right regions.
As should be expected,
an increasing density with an attractive interaction
leads to a lower diffusivity.\cite{Bowker78}
Although the low densities make it difficult
to obtain good statistics,
the diffusivity is clearly reduced by approximately 25\%
from the dilute gas value.

The second parameter which must be determined is $K_i$.
As noted in Sec.~\ref{sec:theory},
the assumption in the macroscopic theory
that $K_i$ is independent of $r$
should be checked for any given system.
Equation~(\ref{eq:Ki-from-J}) shows that $K_i \approx J_a'$.
In Refs.~\onlinecite{Krishnamachari96} and~\onlinecite{McLean96},
we show that $J_a'$
may be measured from simulations
through microscopic considerations,
and that it changes very little with island size.
That is taken as evidence
that the assumption that $K_i$ is independent of $r$
is likely to be satisfied.
However, our attempts at describing the simulated decay
using $K_i$ calculated from $J_a'$ in this way
have so far failed to agree quantitatively
with the observed island decay,
leading us to develop the methods described in the
following paragraphs for determining $K_i$ from purely thermodynamic
arguments.
This procedure below was carried out for only one size island,
under the assumption that $K_i$ is independent of island size.

Our method of numerically extracting
the interface transfer constant $K_i$
is closely related to the shape fluctuation analysis
of Khare {\it et~al.}\cite{Khare96}
They study how the equilibrium fluctuations of the island {\em shape}
depend on the interface transfer constant,
the surface diffusion constant, and the edge diffusion constant;
we will consider
the equilibrium fluctuations of the island {\em size}.
However, we stabilize the island against changes in size
by placing it in a finite box with conserved order parameter,
while Khare {\it et~al.} stabilize the island size
by externally adjusting the chemical potential
under conditions of nonconserved order parameter.
The equations in Ref.~\onlinecite{Khare96}
therefore do not apply directly to our results.

The parameter $K_i$ may be determined through its relation to
the mobility of an interface.
Because a current of adatoms detaching from an interface
causes the interface to move,
$K_i$ is closely related to the mobility for interface motion
under the driving force of a vapor density imbalance.
In principle, this mobility could be directly measured
by putting an interface in interaction
with a vapor of density lower than the equilibrium density.
The interface mobility could then be determined
from the rate at which the interface recedes.
In practice for a computer simulation, however,
it is very difficult to
keep the vapor concentration constant near the interface.
Detaching atoms continually increase the vapor density,
and this source must be counterbalanced by an adatom sink.
However, the required strength of this sink
is not known {\it a priori}.
Indeed, it is even difficult to distinguish whether
a density increase near the interface is due to net detachment
or is due to a fluctuation in the vapor density.

The interface mobility is closely related to the fluctuations
of the interface position in an equilibrium system.
By considering an equilibrium system,
the need for an adatom sink is eliminated.
However, one must be careful because of
the nature of the interface-vapor interaction. 
When an atom detaches from the interface,
the local vapor density is enhanced momentarily.
Because the vapor has a finite diffusivity,
this enhancement will persist for some time,
influencing the subsequent behavior of the interface.
This memory effect means that the interface position
does not perform a random walk,
and the fluctuations of the interface position are thus slowed.
The strength of the memory effect depends on the vapor diffusivity,
and also
on the system size because of the long range nature
of the solution to the 2D diffusion equation.
Note that these considerations do not influence
the average vapor density near the interface.
The memory effect is due to time correlations
between fluctuations of the vapor density
and fluctuations of the interface position.

We have not obtained an analytic form for the
interface fluctuations in the presence of such a vapor memory effect.
However, it is clear that the memory can be removed
by causing the adatom vapor to equilibrate quickly
on the time scale of interface motion.
In simulation, this can be achieved
by randomly redistributing all monomer adatoms to other monomer sites
between every atomic move.
This explicitly does not affect the interface mobility,
because all types of atomic moves which involve the interface are
unaffected;
the adatom redistribution is done in a way such
that it will never cause
the creation of new nearest neighbor bonds.
By thus removing the vapor memory effect,
the interface fluctuations may easily be analyzed to obtain
the interface mobility,
and hence the interface transfer constant $K_i$.

In fact, the vapor memory is not completely removed by this artifice,
because clusters in the vapor are stable.
However, the density of clusters in the vapor
is small in our simulations.
Also, each cluster will have some mean time to dissociation,
so that the memory introduced by clusters of a given size
decays exponentially.
For these reasons, the effect of clusters should be negligible.

In detail then,
we simulate an island in equilibrium with an adatom vapor
in a closed, conserved mass system.
Note that by simulating an island, rather than a straight interface,
any angular dependencies are automatically averaged.
Because mass is conserved, the island size is stabilized at some
equilibrium size $r_{\text{eq}}$.
When the island fluctuates to be either larger or smaller than this,
the resulting change in the surrounding adatom density
tends to restore the island size.
This can be understood in terms of
a thermodynamic potential for the system
which is a function of island radius, $\Phi(r)$.
Reference~\onlinecite{Krishnamachari96} discusses this potential
at length.\cite{potential_note}
The equilibrium radius $r_{\text{eq}}$ is at
the minimum of this potential.

The fluctuations in the island size can be seen
as a random walk of the island radius,
characterized by some interface diffusivity $D_i$.
Note that the diffusivity of the {\em interface} $D_i$
is distinct from the adatom diffusivity $D$.
In general the behavior of the interface would be affected
by the adatom diffusivity $D$
through the memory effect described above.
For these simulations, however, monomer rearrangement is employed
to achieve fast vapor equilibration,
so that the interface diffusivity obtained
is a property of the interface alone.

The random walk of the island radius
is biased by the thermodynamic potential $\Phi(r)$,
so that it tends to return towards $r_{\text{eq}}$.
The strength of this bias is given
by the thermodynamic mobility of the
interface $B_i$,
which is defined by the equation
\begin{eqnarray}
\deriv{r}{t}&=&-B_i \deriv{\Phi}{r} \nonumber \\
    &=&-B_i \left(\left.\deriv{}{\rho}\left(\deriv{\Phi}{r}\right)
		\right|_{\text{eq}}
	(\rho -\rho^{\text{eq}}) +\cdots \right)\;, \label{eq:Bdef}
\end{eqnarray}
In the second equality we have Taylor expanded $d\Phi/dr$ in $\rho$
about the equilibrium density $\rho^{\text{eq}}$,
for reasons which will become clear below.
Note that the interface mobility $B_i$ and the interface diffusivity
may be related through the generalized Einstein relation
$D_i=kT\,B_i$.

\begin{figure}[b]
\centering\leavevmode
\epsfxsize=4in
\epsffile{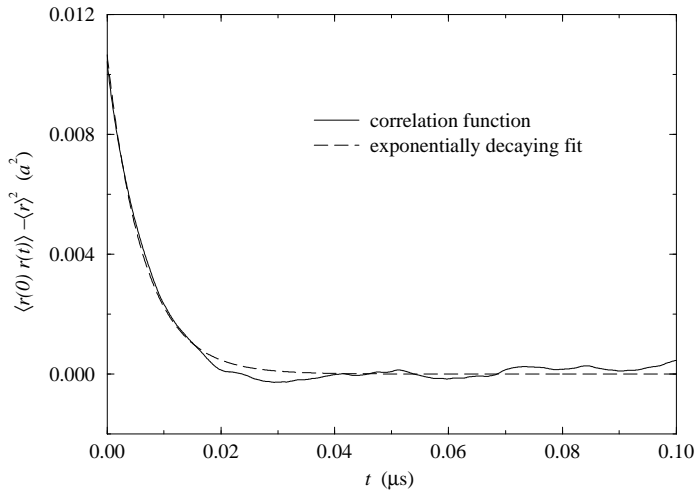}
\caption{The time correlation function for the interface position $r$,
showing the function calculated from one equilibrium simulation
and a least squares fit of the form in Eq.~(\ref{eq:r-corr}).}
\label{fig:autor}
\end{figure}

To characterize the fluctuations of a random walk in a potential,
we approximate the potential as quadratic in $r$:
\begin{equation}
\Phi(r)=\frac{1}{2}c(r-r_{\text{eq}})^2\;,
\qquad c\equiv\left.\frac{d^2\Phi}{dr^2}\right|_{\text{eq}}\;,
\end{equation}
where $c$ gives the curvature of the potential well.
The value of $c$ can be calculated
from the known form\cite{Krishnamachari96} of $\Phi(r)$,
but will not be required for the following analysis.
Note that the Boltzman distribution
in a quadratic potential is Gaussian,
\begin{equation}
\text{Probability}(r) \propto \exp\!\left(\frac{-\Phi(r)}{kT}\right)
=\exp\!\left(\frac{-c(r-r_{\text{eq}})^2}{2kT}\right)\;.
\label{eq:Phiwell}
\end{equation}
The Green's function for diffusive motion
within a quadratic potential is therefore also Gaussian,
with an exponentially relaxing mean and standard deviation:
\begin{mathletters}
\begin{equation}
G(r,t;r_0)=\frac{1}{\sqrt{2\pi}\,\sigma}
	\exp\!\left(-\frac{(r-r_m)^2}{2\sigma^2}\right)\;,
\end{equation}
\begin{equation}
r_m=r_{\text{eq}}+(r_0-r_{\text{eq}}) e^{-B_ict}\;,
\end{equation}
\begin{equation}
\sigma^2=\frac{D_i}{B_ic}\left(1-e^{-2B_ict}\right)\;,
\end{equation}
\end{mathletters}%
where $r_m$ and $\sigma$ are the time dependent mean
and standard deviation of the Gaussian form.
Recall that the Green's function can be interpreted
as the probability
distribution which evolves from
an initial delta function distribution at $r_0$.
It can therefore be used to find the time correlation function
for the interface position $r$,
\begin{equation}
\langle r(0)\,r(t)\rangle -r_{\text{eq}}^2
= \frac{D_i}{B_ic}\:e^{-B_ict}\;,
\label{eq:r-corr}
\end{equation}
the prefactor of which may of course be simplified with the Einstein
relation.
Thus $B_i$ is found by measuring
the time correlation function of the island radius
(using $r\equiv \sqrt{\text{Area}/\pi}$)
and fitting it with Eq.~\ref{eq:r-corr},
as shown in Fig.~\ref{fig:autor}.

The interface transfer constant $K_i$ is then obtained from $B_i$
by comparing the ways they each describe interface motion.
$K_i$ describes how an interface moves
due to a vapor density imbalance
[see Eq.~(\ref{eq:simpleIT})],
\begin{equation}
-\deriv{r}{t}\frac{2\pi r}{\Omega}=I_i=
	2\pi r K_i(\rho^{\text{eq}}-\rho(r))\;.
\end{equation}
In the Taylor expansion of Eq.~(\ref{eq:Bdef}),
the interface mobility was also described in terms of densities,
so that by comparing these equations we obtain
\begin{equation}
K_i=-\frac{1}{\Omega} B_i
\left.\frac{d^2\Phi}{dr\,d\rho}\right|_{\text{eq}}
=\frac{1}{\Omega} B_i c
\left(-\left.\deriv{r}{\rho}\right|_{\text{eq}}\right)\;.
\label{eq:KfromB}
\end{equation}
The derivative $dr/d\rho$ is given by purely geometrical
considerations.
Note that it is not necessary to know $c$ in order to obtain
$K_i$ with this method;
the product $B_ic$ is obtained directly
from the decay length of Eq.~(\ref{eq:r-corr}).
As a consistency check,
$c$ was calculated from $\Phi$
as known from the equation of state for
the Ising model.\cite{Krishnamachari96}
It was then used to check the initial amplitude
of the time correlation function.

\section{Quantitative evaluation of decay simulations}
\label{sec:compare}
The above procedures yield values of
$K_i=(2.47\pm 0.19)\times 10^{10}\,a/\text{s}$
and, at the equilibrium concentration for the initial island,
$D=(1.75\pm 0.2)\times 10^9\,a^2/\text{s}$.
Using Eq.~(\ref{eq:rtranspit}),
these values together with the average pit radius of $R=38.5$
yield transition radii of $r'_t=R/4550\approx8\times10^{-3}\,a$ and
$r'_t=R/1.0037\approx38\,a$.
Thus the island, which is initially of size $r=15\,a$,
remains well within the diffusion limited regime
throughout the decay.
Figure~\ref{fig:Cu-decay} shows the results
of numerically integrating
Eq.~(\ref{eq:simcurr}) using these parameters.
The agreement with the simulation results is very good.
As expected based on the transition radii,
the predicted decay is extremely insensitive to the value of
the interface transfer constant.
The deviation at very small islands may be due either to a breakdown
in the phenomenological expression of the modified GT equation,
or to a breakdown in the assumption
that $K_i$ is size independent.
A breakdown of the quasi-steady-state assumption seems unlikely,
since even at the highest recorded interface velocities
each monomer makes over 1000 moves
in the time that the interface recedes by one lattice spacing.

In order to test the macroscopic theory in a regime where
the interface transfer process is important,
a second set of ten decay simulations
were run in which the diffusion rate was greatly increased.
This was achieved by reducing the barrier for moves
from 0-fold to 0-fold coordination.
That barrier was changed
from $0.697$\,eV to $0.0$\,eV.
This unphysical sounding ``barrier'' is possible
because our simulation uses the barrier parameters
only to determine rates.
Note that at room temperature the same change in the rate
of these moves
could be achieved with a barrier reduction of only $0.14\,\text{eV}$.
The large magnitude of the barrier change,
required to obtain the following results,
is necessary here because of the high temperature of the simulation.

\begin{figure}[b]
\centering\leavevmode
\epsfxsize=4in
\epsffile{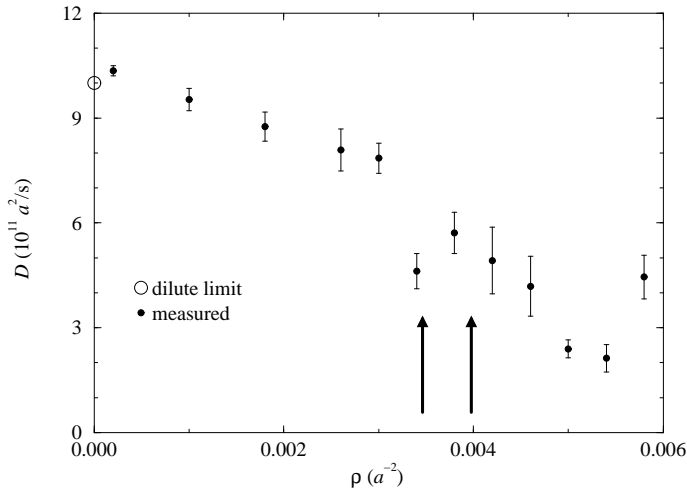}
\caption{The dependence of the diffusion coefficient on adatom density
in the modified simulation which has a faster diffusion rate,
obtained from direct measurements
of current densities under enforced concentration gradients.
Open circle shows the theoretical dilute limit value.
The arrows indicate the concentration of vapor adatoms required
for equilibrium with
each of the pit (left arrow) and the island (right arrow)
at the beginning of the decay simulation.}
\label{fig:CuDd=0}
\end{figure}

As determined by the methods described above,
the diffusivity is even more concentration dependent
than in the previous case.
Figure~\ref{fig:CuDd=0} shows that
at the vapor density which is in equilibrium with the initial island,
the diffusivity $D=5.0\times\!10^{11}\,a^2/\text{s}$
is one half of the dilute limit value of
$D=1.0\times\!10^{12}\,a^2/\text{s}$.
As the island shrinks, the concentration increases
and the diffusion coefficient decreases significantly.
If we are to describe the entire decay with a single effective
diffusion coefficient,
it will be yet lower than $5.0\times\!10^{11}\,a^2/\text{s}$.

Since non-diffusion barriers were unchanged,
the same value of $K_i$ as above applies here.
Using these values in Eq.~(\ref{eq:rtranspit}),
the left hand side is $D/(K_iR) = 0.317$.
As a result, Eq.~(\ref{eq:rtranspit}) has no roots,
and the interface transfer process is the most important process
in limiting the decay rate during the entire island decay.
Note however that this decay is not far
into the interface-transfer limited
regime, so that the diffusion coefficient is still expected to be of
importance.

\begin{figure}[b]
\centering\leavevmode
\epsfxsize=4in
\epsffile{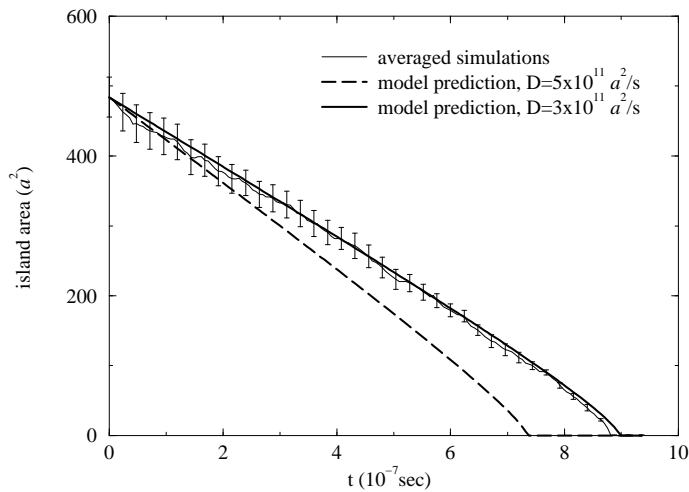}
\caption{Simulated island decay with fast diffusion,
showing the averaged island area versus time,
referenced to time of disappearance.
Error bars represent standard deviations of the mean.}
\label{fig:D0decay}
\end{figure}

Shown in Fig.~\ref{fig:D0decay} is the averaged decay
from simulation,
as well as analytic predictions.
In this case the analytic results are sensitive
to both $D$ and $K_i$ as expected.
Note that the decay rate is two orders of magnitude larger
than in the previous case.
The relationship of interface velocity and diffusivity
is similar to the previous case,
supporting the use of the quasi-steady-state assumption.
The dashed curve,
which uses the value $D=5.0\times\!10^{11}\,a^2/\text{s}$,
qualitatively succeeds in showing this increase in decay rate,
but deviates as the island becomes smaller.
The entire decay can be matched quite well by using the value
$D=3.0\times\!10^{11}\,a^2/\text{s}$,
as shown by the solid curve.
This can be partially justified by the fact that the
smaller island results in higher adatom concentrations,
and hence a lower diffusion coefficient.
However, because of uncertainties in the calculations
of $D$ and $K_i$,
it is difficult to compare simulation and analytic theory closely.

In Sec.~\ref{sec:limit} we saw that in treatments of systems
with no clear outer boundary, under certain approximations,
the diffusion limited and interface-transfer limited regimes
are characterized by power law decay of the island area,
with respective exponents of $2/3$ and 1.
For these simulations,
there is a definite outer boundary which is near the island
and the various approximations are not valid,
so that power law behavior cannot be achieved.
However, at the end of the decay in Fig.~\ref{fig:Cu-decay}
the decay rate increases markedly,
whereas in Fig.~\ref{fig:D0decay} the increase in decay rate is less
pronounced.
In this sense, the diffusion limited decay is ``less linear''
than the decay dominated by the interface transfer process.
This distinctive feature is correctly reflected by the analytic
predictions.

In principle, simulations might be run
further into the interface-transfer
limited regime by further increasing the difference between
the energy barriers for interface
transfer and for diffusion.
However such an approach is impractical,
as the program must spend most of its time
processing diffusion events,
and therefore decay occurs very slowly in real computer time.
Other, more artificial means
of increasing the effective diffusion rate,
such as the monomer rearrangement scheme
described in Sec.~\ref{sec:macroparams},
might circumvent this difficulty.

\section{Conclusion}
Isolated features on a surface will tend to decay in size due to the
Gibbs-Thomson effect.
This decay implies that there is mass flow in the adatom vapor
surrounding the island,
flowing from the island to the surrounding environment.
There are three processes in this mass flow to be considered:
interface transfer at the island edge,
diffusion across the surface,
and incorporation into the environment at some outer boundary.
For various surface parameters and various surface configurations,
each of these processes may be the rate limiting one,
so that there are three general cases of the island decay.

With certain approximations,
the limiting regimes of these cases
lead to a power law dependence of the island size on time.
For the interface-transfer limited regime,
this leads to the island area decaying as $A\propto(t_0-t)$.
For the outer-boundary limited regime,
this leads to the island area decaying as $A\propto(t_0-t)^{2/3}$.
For the diffusion limited regime, no power law is strictly valid,
but under a screening length hypothesis
the $A\propto(t_0-t)^{2/3}$ power law can be argued.

The existence of these power laws
may make it possible to experimentally
distinguish between the regimes
by measuring exponents for observed feature decay.
Both the outer-boundary limited regime
and the diffusion limited regime
could probably be observed on physical surfaces.
The interface transfer regime is unlikely to occur
on clean metal surfaces,
but may be important on other surfaces
or on metal surfaces covered with adsorbates.
It should be noted, however,
that it can be quite difficult to distinguish
between the two power laws based on a single decay.
Data must either have very low uncertainty in measured island areas,
or consist of many independent decays
(as in Ref.~\onlinecite{Morgenstern96}),
in order to be a good test of the exponent.

Computer simulations based on the solid-on-solid model and simple
energetics exhibit the island decay behavior.
Comparison of simulations and the analytic theory demonstrate
that this theory correctly captures the essential elements of
island decay.
Although some limiting regimes of the theory
cannot be fully achieved in simulation,
the simulations match the theoretical predictions well
for an important range of surface parameters.
Note that this is the case in spite of the fact that the theory is
thermodynamic and macroscopic in nature,
while the simulation involves features of atomic scale.
We conclude that even at atomic scales
the analytic theory includes all
the physics necessary to describe
the behavior of a surface governed by simple energetics.

The relationship between the parameters
governing the microscopic dynamics
of the system (e.g., the energy barriers for atomic moves)
and the macroscopic dynamics parameters $D$ and $K_i$
is quite complex.
We have shown methods of obtaining the macroscopic parameters
from measurements of the simulation behavior.
Further investigations of the relationship between microscopic and
macroscopic parameters would be of interest,
since in principle the macroscopic behavior is completely determined
by the microscopic characteristics.

\acknowledgments
The authors wish to thank Leonard Feldman,
Robert Silsbee,
Eugene Kolomeisky,
Chris Henley,
Gerard Barkema,
Tatjana \'Cur\v{c}i\'c,
Jack Blakely,
and Georg Rosenfeld
for valuable conversations.
We particularly thank Rein Breeman for providing
energy barriers for atomic moves,
upon which our simulation parameters were based.
This work was generously supported by the National Science Foundation
through the Cornell Materials Science Center
(NSF-DMR-9121654),			
and through grants NSF-DMR-9313818,		
NSF-GER-9022961, and NSF-DMR-9200469,		
and by the Air Force Office of Scientific Research
(AFOSR-91-0137					
and AASERT number F49620-93-1-0504).		
Early portions of this work were performed
at Sandia National Laboratories
supported by the US Department of Energy
under contract DE-AC04-94AL85000.

\bibliographystyle{prsty}
\bibliography{journals,endnotes,GT,equipment,materials,diffusion,%
		Ising,growth,coarsening,comp_meth,decay,theses}

\begin{thebibliography}{10}
\bibitem{GTisKelvin}
This form of the Gibbs-Thomson relation is sometimes referred to as the Kelvin
  equation, especially in the context of adsorption in porous media.
\bibitem{potential_note}
The thermodynamic potential is described in detail in
  Ref.~\onlinecite{Krishnamachari96}, where it is given the symbol $G$; see
  Eq.~(16) and Figs.~2 and~6 in that reference.
\end{thebibliography}
\end{document}